\documentclass[pre,letterpaper,floatfix,superscriptaddress,11pt]{revtex4}
\usepackage{graphicx}
\begin{document}

\title{An Accelerated Multiboson Algorithm for Coulomb Gases with Dynamical Dielectric Effects} 

\author{A.~Duncan}
\affiliation{Department of Physics and Astronomy\\
University of Pittsburgh, Pittsburgh, PA 15260}
\author{R.D.~Sedgewick}
\affiliation {Department of Biological Sciences \\
Carnegie Mellon University \\
Pittsburgh, PA 15213}


\begin{abstract}

A recent reformulation \cite{ourmultibos} of the problem of Coulomb
gases in the presence of a dynamical dielectric medium showed that
finite temperature simulations of such systems can be accomplished on
the basis of completely local Hamiltonians on a spatial lattice by
including additional bosonic fields.  For large systems, the Monte
Carlo algorithm proposed in Ref.~\cite{ourmultibos} becomes inefficient due
to a low acceptance rate for particle moves in a fixed background
multiboson field. We show here how this problem can be circumvented by
use of a coupled particle-multiboson update procedure that improves
acceptance rates on large lattices by orders of magnitude. The method
is tested on a one-component plasma with neutral dielectric particles
for a variety of system sizes.

\end{abstract}

\maketitle

\section{Introduction}

Some years ago, Maggs and collaborators \cite{maggs:prl} introduced a
local reformulation of the statistical mechanics of a Coulomb gas of
mobile charged entities, relying on the fact that the long-range
Coulomb interaction could be recovered from a Hamiltonian written in
terms of an electric field variable coupled to the charge density via
Gauss' Law. Since the publication of this paper, more recent work has
focused on improving the efficiency and capability of the method so
that it may be applied to larger and more realistic systems
\cite{maggs:alone,maggs:worm,maggs:trail,ourpaper:fft,ourpaper}. The
partition function in this approach is written as a functional
integral over an electric field variable containing a physical
longitudinal (gradient) part incorporating the electrostatic potential
energy, as well as an unphysical transverse (curl) part.  The latter
component of the electric field can easily be seen to decouple from
the charged particle motion, \textit{provided the dielectric constant
does not change in the course of the simulation} (i.e. is
non-dynamical). Note that a spatially varying dielectric medium is
perfectly allowable in the context of the original method: the
integration over the transverse parts of the electric field variable
only introduces spurious interactions if the dielectric function is
also a function of the particle locations. This latter situation is of
course quite common in simulations of biophysical interest, for
example if the charges are attached to a polymer undergoing
conformational changes.

In a recent paper \cite{ourmultibos}, we suggested a new algorithm
for compensating for the spurious interactions introduced by the
integration over the curl part of the electric field in situations of
this type. The extra terms actually correspond simply to the inverse
determinant of the Poisson operator
$\vec{\nabla}\cdot(\epsilon(\vec{r})\vec{\nabla})$. Removing them
therefore amounts to the problem of introducing a positive power of
the determinant of a local operator into a functional integral. This
problem has been intensively studied in elementary particle theory,
where the inclusion of virtual quark effects in lattice quantum
chromodynamics simulations require the introduction of just such a
determinant in a functional integral. In particular, the multiboson
method introduced by L\"uscher \cite{luescher} seems ideally suited
for the task at hand, for reasons explained below.  Re-expressing the
Poisson determinant in terms of an integral over a finite number $N_B$
of additional bosonic fields, we arrive at a local Hamiltonian and a
partition function that can be easily simulated by standard
Metropolis/heatbath techniques. In this approach, an update of the
system proceeds by successive, and independent, updates of (a) charged
particle locations, (b) electric field values (on the links of the
lattice), and (c) all $N_B$ multiboson fields (on the sites of the
lattice).

The algorithm studied in Ref.~\cite{ourmultibos} does suffer however
from a serious drawback. The multiboson fields are in general rather
strongly coupled to the particle locations, so that a particle move
leaving the boson fields unchanged becomes difficult when it induces a
large corresponding change in the multiboson energy, without allowing
the multiboson fields to relax in response to the new particle
location. The problem becomes acute when the number $N_B$ of
multiboson fields is increased: for example, to model more accurately 
the low momentum modes on a large lattice. One then finds an
acceptance rate for particle moves that decreases exponentially with
$N_B$. The goal of the present paper is to show that a modified
update procedure which couples particle moves and boson field updates
greatly ameliorates this acceptance problem. Efficient simulation with
a large number of multiboson fields is necessary for performing
realistic simulations on large lattices.

In Section 2, we review the basic multiboson algorithm introduced in
Ref.~\cite{ourmultibos}, valid for systems with dynamically varying
dielectric functions. The new coupled update method, allowing the
multiboson fields to relax appropriately as the particles are moved,
is explained in detail in Section 3. Detailed numerical tests of the
new method, with comparisons to the original, uncoupled approach, are
presented in Section 4. Finally, we offer some concluding remarks in
Section 5.

\section{Multiboson Formulation of Coulomb Gas Systems with Varying Dielectric}
\label{sec:theproblem}

In this section we review the multiboson reformulation of
Ref.~\cite{ourmultibos} for Coulomb gas systems with varying dielectric.
Consider the partition function of a system consisting of $N$ free
charges (mobile or fixed) $e_i$ at locations $\vec{r}_{i}$,
corresponding to a free charge density
\begin{equation}
\label{eq:chargedens}
   \rho(\vec{r}) = \sum_{i}e_{i}\delta(\vec{r}-\vec{r}_{i})
\end{equation}
where the system is also described by a linear dielectric function
(here assumed isotropic) $\epsilon(\vec{r})$.  We shall suppress the
dependence of $\epsilon(\vec{r})$ on particle locations, but the procedure 
we shall describe is precisely devised to properly account for this 
dependence in the electrostatic energetics of the problem.
Breaking the electric displacement into longitudinal and
transverse parts using the
general Helmholtz decomposition,
\begin{eqnarray}
\label{eq:helmdecomp1}
  \vec{D}(\vec{r}) &=& -\epsilon(\vec{r})\vec{\nabla}\phi(\vec{r})+\vec{\nabla}\times\vec{A}(\vec{r}) \\
  \label{eq:helmdecomp2}
                 &=& \vec{D}^{||}(\vec{r})+\vec{D}^{\rm tr}(\vec{r}).
\end{eqnarray}
As the transverse and longitudinal components are orthogonal to each
other, one has
\begin{equation}
\label{eq:decouple}
  \int d\vec{r}\frac{\vec{D}^{2}}{\epsilon(\vec{r})}=\int d\vec{r}\frac{\vec{D}^{||}(\vec{r})^{2}}{\epsilon(\vec{r})}
  +\int d\vec{r}\frac{\vec{D}^{\rm tr}(\vec{r})^{2}}{\epsilon(\vec{r})}.
\end{equation}
It follows that the imposition of Maxwell's second law (curl-free
electric field) implies that the electrostatic energy of the system is
given purely in terms of the longitudinal part of the electric
displacement
\begin{equation}
\label{eq:eenergy}
 H_{\rm es} = \frac{1}{2}\int d\vec{r}\, \frac{\vec{D}^{||}(\vec{r})^{2}}{\epsilon(\vec{r})}
 \end{equation}
Accordingly, the canonical partition function for the system at
inverse temperature $\beta$ becomes
\begin{equation}
\label{eq:partfunc}
  Z=\int\prod_{i=1}^{N}d\vec{r}_{i}e^{-\beta H_{\rm es}}
\end{equation}
where $\vec{D}^{||}$ must be determined by first solving
$-\vec{\nabla}\cdot(\epsilon\vec{\nabla}\phi)=\rho$, from which one
obtains
$\vec{D}^{||}=-\epsilon(\vec{r})\vec{\nabla}\phi(\vec{r})$. Note that
the (non-existent) transverse part of the displacement field plays no
role in this result.

In Ref.~\cite{ourmultibos} we have shown how to write a representation for
the above partition function by writing the Coulomb energy Boltzmann
weight $e^{-\beta H_{\rm es}}$ in terms of an integral over an
unrestricted electric displacement field $\vec{D}(\vec{r})$ (i.e. one
containing both longitudinal and transverse parts), as well as over a
set of multiboson fields with a local Hamiltonian. The integration
over the multiboson fields is arranged to generate a factor which
exactly cancels the contribution from the integral over the spurious
transverse part $\vec{D}^{\rm tr}(\vec{r})$ for arbitrary dielectric
functions $\epsilon(\vec{r})$. Although the factor itself is a
determinant and highly non-local, the multiboson Hamiltonian is local
and therefore conventional Monte Carlo simulation techniques can be
applied.

First, we review some notational issues (for further details, see
\cite{ourmultibos}).  In order to write a well-defined functional
integral for $Z$, we consider a lattice Coulomb gas with an electric
displacement vector field $D_{n\mu}$ defined on lattice links, where
lattice sites are denoted $n$ and $\mu=1,2,3$ indicating the spatial
direction of the link. Dielectric values are associated with links on
the lattice (rather than sites) so that the dielectric function on
the lattice becomes $\epsilon_{n\mu}$.  Introducing left
(resp. right) lattice derivatives $\bar{\Delta}$ (resp. $\Delta$),
the lattice version of Eq.~\ref{eq:helmdecomp1} becomes
\begin{eqnarray}
\label{eq:helmlat}
  D_{n\mu} &=& D_{n\mu}^{||}+D_{n\mu}^{\rm tr}  \\
  &=& -\epsilon_{n\mu}\Delta_{\mu}\phi_{n} + \sum_{\nu \sigma} \epsilon_{\mu\nu\sigma}\bar{\Delta}_{\nu}A_{n\sigma}
\end{eqnarray}
Here the electrostatic potential is represented by a lattice site
field $\phi_{n}$ satisfying the Poisson equation
\begin{eqnarray}
\label{eq:lattPoiss}
 -\sum_{\mu}\bar{\Delta}_{\mu}(\epsilon_{n\mu}\Delta_{\mu}\phi_{n}) &=& \rho_{n} \\
\label{eq:lattdens}
 \rho_{n} &=& \sum_{i} e_{i}\delta_{nr_{i}}
 \end{eqnarray}
It is then straightforward to show that the contribution to the functional integral
\begin{eqnarray}
 Z^{\prime}\equiv \int \prod_{n\mu}dD_{n\mu}e^{-\frac{\beta}{2}\sum_{n\mu}D_{n\mu}^{2}}
\end{eqnarray}
from the integration over the unphysical transverse degrees of freedom
$D_{n\mu}^{\rm tr}$ yields the factor
$(\prod_{n\mu}\sqrt{\epsilon_{n\mu}})\;{\rm det}^{-\frac{1}{2}}({\cal
M})$, where ${\cal M}$ is essentially the lattice Poisson operator:
\begin{equation}
 \label{eq:defop}
  {\cal
    M}\lambda_{n}\equiv(-\sum_{\mu}\bar{\Delta}_{\mu}\epsilon_{\mu}\Delta_{\mu})\lambda_{n}
=\sum_{i=1}^{6}\epsilon_{ni}\lambda_{n}-\sum_{i=1}^{6}
  \epsilon_{ni}\lambda_{n+i}
\end{equation}
The unwanted determinant factor induced by the transverse
integrations can be removed by introducing a set of additional
fields with a local Hamiltonian chosen to generate a \textit{positive}
square-root determinant canceling the unwanted contribution.  As
explained in \cite{ourmultibos}, one begins from a uniform
polynomial approximation to the function $1/s$ in the interval
[$\delta,1$] for small $\delta$. In terms of the complex roots of
the polynomial $z_k=\mu_k+i\nu_k$, one may choose, for example (the
choice is not unique and partly a matter of convenience
\cite{luescher}) the Chebyshev polynomial of order $2N_B$ with
\begin{eqnarray}
 \label{eq:chebpoly}
 \frac{1}{s} &\simeq&  P(s) \equiv  C\prod_{k=1}^{N_B}((s-\mu_{k})^{2}+\nu_{k}^{2})  \\
  \mu_{k} &=& \frac{1}{2}(1+\delta)(1-\cos{\frac{2\pi k}{2N_B+1}}) \\
  \nu_{k} &=& \sqrt{\delta}\sin{\frac{2\pi k}{2N_B+1}} 
\end{eqnarray}
This representation extends as follows to the
determinant of a real positive symmetric operator with spectrum in the
interval [$0,1$]
\begin{equation}
\label{eq:detpoly}
  {\rm det}^{+\frac{1}{2}}({\cal M}) \simeq \prod_{k=1}^{N_B} {\rm det}^{-\frac{1}{2}} (({\cal M}-\mu_{k})^{2}+\nu_{k}^{2})
\end{equation}
where the representation becomes exact in the limit
$N_B\rightarrow\infty,\delta\rightarrow 0$. (In practice, the Poisson
operator needs to be rescaled, so that its spectrum fits in the unit
interval.) The crucial step in the multiboson procedure is the
replacement of the determinant factor in Eq.~\ref{eq:detpoly} by an
equivalent Gaussian integral over a set of $N_{B}$ boson site fields
$\phi^{(k)}_{n},k=1,2,..N_{B}$. The choice of $N_{B}$ is dictated by
the necessity to adequately describe the low eigenvalues of the
Poisson operator ${\cal M}$: in Ref.~\cite{ourmultibos}, this was
determined empirically by studying a model in which the structure
factor of the system was analytically known in the charge-free
limit. More generally, we expect that as the lowest eigenvalue of
${\cal M}$ on a lattice of dimension $L$x$L$x$L$ is typically of order
$1/L^{2}$, the value of $N_{B}$ will have to increase as the lattice
size does.  However, the precise character of the scaling of $N_{B}$
with $L$ clearly depends on the sensitivity of the observables studied
to the infrared spectrum of ${\cal M}$.

Inserting a delta-function to enforce the lattice version of Gauss'
Law (thereby fixing the longitudinal part of the electric
displacement appropriately), the expression found for the correct
partition function (Eq.~\ref{eq:partfunc}) \cite{ourmultibos} becomes

\begin{eqnarray}
\label{eq:mbosaction}
   Z&=&
   \int\prod_{i}d\vec{r}_{i}\prod_{n\mu}dD_{n\mu}\prod_{kn}d\phi^{(k)}_{n}
    \,\delta(\sum_{\mu}\bar{\Delta}_{\mu}D_{n\mu}-\rho_{n})\nonumber \\
  &&\times e^{
-\frac{1}{2}\sum_{n\mu}\log{(\epsilon_{n\mu}})} \,
 e^{-\frac{\beta}{2}\sum_{n\mu}D_{n\mu}^{2}/\epsilon_{n\mu}}\nonumber \\
  \label{eq:partfinal2}
  &&\times e^{-\sum_{k=1}^{N_B}\phi^{(k)}(({\cal M}-\mu_{k})^{2}+\nu_{k}^{2})\phi^{(k)}}
\end{eqnarray}
Note that the effective Hamiltonian appearing in the exponent here is
completely local.  Monte Carlo simulation of the system is in
principle straightforward: a state of the system is characterized by
(i) particle locations $\vec{r}_{i}$, which in general will influence
the dielectric function $\epsilon(\vec{r})$ (which more accurately
should be denoted $\epsilon(\vec{r};\vec{r}_{i})$), (ii) the lattice
displacement field $D_{n\mu}$ (respecting locally the Gauss' Law
constraint), and (iii) the $N_B$ auxiliary scalar fields $\phi^{(k)}$,
and simulation requires updates of all of these variables with a
procedure respecting detailed balance and according to the indicated
Boltzmann weight.

\section{Efficient Monte Carlo Procedure for Coulomb Gas/Multiboson System}

 In \cite{ourmultibos}, the multiboson approach was tested on a system
 of particles (both neutral and charged) with a dielectric different
 from that of the ambient medium.  This model, first studied in
 Ref.~\cite{maggs:alone}, provides a clean testbed for algorithms
 dealing with the full electrostatic energetics of a system with a
 dynamical dielectric function. The latter changes through the
 simulation as the dielectric constant along a link in the $\mu$
 direction from a lattice site $n$ is defined through the relation
\begin{equation}
\label{eq:dieleq}
\frac{2}{\epsilon_{n\mu}} = \frac{1}{\epsilon_n} + \frac{1}{\epsilon_{n+\mu}},
\end{equation}
where $\epsilon_n$ and $\epsilon_{n+\mu}$ are either the background
dielectric constant or the particle dielectric constant depending on
whether there is a particle on site $n$ or site $n+\mu$
respectively. For example, in the simulations discussed below, the
background dielectric constant is set to $1.0$, while the dielectric
constant associated with the particle is set to $0.05$. If the
particles are electrically neutral, then we expect the electric
displacement to vanish \textit{irrespective} of the particle locations,
leading to a flat (wave-number independent) density-density structure
factor.  The simulations of Eq.~\ref{eq:mbosaction} in
Ref.~\cite{ourmultibos} show that with a relatively small number of
multiboson fields ($N_{B}$=4 on a 16x16x16 lattice) the neutral system
yields a flat structure factor within statistical errors. For the rest
of this paper, we shall use this model system to discuss issues of
acceptance rates and simulation efficiency in the multiboson approach.

Unfortunately, the simulation procedure of Ref.~\cite{ourmultibos}, which
we now review briefly, leads to an acceptance rate for particle moves
that falls extremely rapidly as the number of multiboson fields is
increased.  For example, in our 16x16x16 test lattice the procedure of
Ref.~\cite{ourmultibos} gives a particle move acceptance rate of 0.011 when
8 multiboson fields are used, but the acceptance rate drops to less
than 2e-7 when 32 multiboson fields are used.  As we emphasized above,
increasing $N_{B}$ will be necessary if we wish to go to much larger
lattices, in order to apply the method to systems of real biophysical
interest, for example. Specifically, the Monte Carlo procedure of
\cite{ourmultibos} consisted of
\begin{enumerate}
  \item A Metropolis move of a particle (randomly chosen) to a neighboring site,
  holding the electric displacement field $D_{n\mu}$ and the multiboson fields
  $\phi^{(k)}_{n}$ fixed.
  \item A heat-bath (or worm \cite{alet:orig_worm}) update of the electric displacement field $D_{n\mu}$,
  with particle locations and multiboson fields held fixed.
  \item A Metropolis (or heat-bath) update of the multiboson fields, with electric
  displacement field and particle locations held fixed.
\end{enumerate}
The acceptance problems of the method arise from the first step above:
Metropolis moves of the particles along lattice links. As the number
of multiboson fields increases, the change in the contribution of the
multiboson Hamiltonian to the Boltzmann weight increases (roughly
linearly in $N_{B}$) and the acceptance rate for moves decreases
exponentially. The origin of the decrease is simply that the
displacement of a particle to a neighboring site leaving the
multiboson fields unchanged places the system on the tail of the old
multiboson Gaussian, with the multiboson Gaussian corresponding to the
new particle location having very little overlap with the old one. The
heat-bath approaches used in the second and third steps above have of
course no such acceptance problems.
  
The displacement of the multiboson Gaussian induced by a particle move
holds the key to finding a more efficient move algorithm: one clearly
needs to update the particle positions and multiboson fields
\textit{simultaneously}.  We now show that the Gaussian dependence of
the Hamiltonian on the multiboson fields allows us to do this with a
heat-bath approach that yields greatly increased (for larger values
of $N_B$, by several orders of magnitude) acceptance rates for
particle moves. Let us denote the multiboson contribution to the
Boltzmann weight in Eq.(\ref{eq:mbosaction}) by $e^{-S_{\rm mbos}}$.
Now consider a particle move across a link connecting two nearest
neighbor sites $a,b$ on the lattice (i.e. we assume that initially
there is a particle at one, but not both of these sites). The complete
set of multiboson fields on the lattice can be grouped as
$(\phi_{a}^{i},\phi_{b}^{i},\phi^{\prime\;i}),i=1,2,..N_B$, where the
fields \textit{not} on sites $a$ or $b$ are collectively labeled
$\phi^{\prime}$. The next step is to expose the dependence of the
multiboson energy on just the fields at sites $a$ and $b$: this will
allow us to derive a heat-bath procedure for simultaneously moving the
particle between $a$ and $b$ and updating all multiboson fields on
these two sites, thereby ``adapting" the multiboson fields dynamically
to the updated particle position. As the multiboson energy is a
quadratic function of the multiboson fields, we must have
\begin{equation}
  \label{eq:sab}
  S_{\rm mbos} = \sum_{i=1}^{N_B}(A_{i}(\phi_{a}^{i})^{2}+B_{i}(\phi_{b}^{i})^{2}+2G_{i}\phi_{a}^{i}\phi_{b}^{i}
    +C_{i}\phi_{a}^{i}+D_{i}\phi_{b}^{i}) +S_{\rm non-ab}(\phi^{\prime},\epsilon)
\end{equation}
Here, the coefficients $A_i,B_i,G_i,C_i,D_i$, as well as the residual
term $S_{\rm non-ab}$ all depend on the fields $\phi^{\prime}$
\textit{not} on the pair of sites $a,b$, as well as the dielectric
field $\epsilon$, which of course depends on particle locations via
Eq.(\ref{eq:dieleq}). It is straightforward to find explicit
expressions for these coefficients in terms of $\epsilon$ and the
$\phi^{\prime}$ fields. We shall use Greek indices
$\rho,\sigma,\tau,..$, running over six values, to denote positive and
negative spatial directions out of a given lattice site, while unit
vectors in the corresponding directions are denoted $\hat{\rho}$
etc. Now let us assume that site $b$ is the nearest neighbor of site
$a$ in the $\sigma$ direction. Defining the site field
\begin{equation}
   E_{n}\equiv \sum_{\rho}\epsilon_{n\rho}
   \end{equation}
   one finds for the quadratic coefficients
   \begin{eqnarray}
   A_{i} &=& (E_{a}-\mu_{i})^{2}+\sum_{\rho}\epsilon_{a\rho}^{2}+\nu_{i}^{2} \\
   B_{i} &=& (E_{b}-\mu_{i})^{2}+\sum_{\rho}\epsilon_{b\rho}^{2}+\nu_{i}^{2} \\
   G_{i} &=& -\epsilon_{a\sigma}(E_{a}+E_{b}-2\mu_{i})
   \end{eqnarray}
   while for the coefficient linear in $\phi_{a}$ one has
   \begin{equation}
   C_{i}=-2\sum_{\rho\neq\sigma}(E_{a}+E_{a+\hat{\rho}}-2\mu_{i})\epsilon_{a\rho}\phi^{i}_{a+\hat{\rho}}
    +2\sum_{\rho+\tau\neq 0}\epsilon_{a\rho}\epsilon_{a+\hat{\rho},\tau}\phi^{i}_{a+\hat{\rho}+\hat{\tau}}
    \end{equation}
with a similar equation (replacing $a$ by $b$) for the coefficient $D_{i}$.
     
Defining diagonal matrices ${\cal A}_{ij}=A_{i}\delta_{ij}$,
${\cal B}_{ij}=B_{i}\delta_{ij}, {\cal G}_{ij}=G_{i}\delta_{ij}$, and 
a $(2N_{B})$x$(2N_{B})$ matrix $M$ by\\
     \[ M= \left(  \begin{array}{cc}
           {\cal A} &  {\cal G}         \\
           {\cal G}  & {\cal B}     
           \end{array} \right)       \]
and a $2N_{B}$ dimensional column vector containing the coefficients
$C_{i},D_{i}$:
       \[  V= \left( \begin{array}{c}
            C \\ D    
            \end{array} \right) \] \\
with a similar representation for the fields $\phi_{a},\phi_{b}$:
     \[  \Phi= \left( \begin{array}{c}
            \phi_{a} \\ \phi_{b}    
            \end{array} \right) \]     
the Boltzmann weight for the fields $\phi_{a},\phi_{b}$   may be written
\begin{eqnarray}
      e^{-\Phi M\Phi-V\Phi} &=& e^{-(\Phi+\frac{1}{2}VM^{-1})M(\Phi+\frac{1}{2}M^{-1}V)+\frac{1}{4}VM^{-1}V} \\
      &=& e^{-\Psi M\Psi}e^{\frac{1}{4}VM^{-1}V}
\end{eqnarray}
with
\begin{equation}
     \Psi \equiv \Phi+\frac{1}{2}M^{-1}V
\end{equation}
As the multiboson fields for different index $i$ do not interact, the
$(2N_B)$x$(2N_B)$ matrix algebra implicit in the above equations
actually decouples into $N_B$ independent 2x2 problems. In particular,
the total weight for the $\phi_{a},\phi_{b}$ field pair, with a
particle at either $a$ or $b$, is found by integrating out these
fields:
\begin{eqnarray}
      \label{eq:intwgt}
      W_{ab} &\equiv& \int d\Phi e^{-\Phi M\Phi-V\Phi} \\
        &=& {\rm det}^{-\frac{1}{2}}(M)e^{\frac{1}{4}VM^{-1}V} \\
        &=& \prod_{i}(A_{i}B_{i}-G_{i}^{2})^{-\frac{1}{2}}e^{\frac{1}{4}(B_{i}C_{i}^{2}+A_{i}D_{i}^{2}
        -2G_{i}C_{i}D_{i})/(A_{i}B_{i}-G_{i}^{2})}
        \end{eqnarray}
      
With these preliminaries, we can now state the procedure needed
to implement a coupled heat-bath particle move/multiboson field
update. The first step is to determine a relative Boltzmann
weight for placing a particle either at site $a$ or site
$b$. This weight is determined as a product of three factors:
\begin{enumerate}
\item If the particle is charged, a contribution from the
electric field term
$e^{-\frac{\beta}{2}\sum_{n\mu}^{\prime}D_{n\mu}^{2}/\epsilon_{n\mu}}$,
where the prime indicates that only lattice links connected to
site $a$ \textit{or} site $b$ are included. Here the dependence on
particle location is entirely through the change a particle move
induces in the dielectric field.
\item A contribution from the dielectric factor
$e^{-\frac{1}{2}\sum^{\prime}_{n\mu}\log{(\epsilon_{n\mu}})}$,
with the same interpretation of the primed sum.
\item A contribution from the multiboson energy
$W_{ab}e^{-S_{\rm non-ab}}$ arising from the total integrated
weight $W_{ab}$ of the $\phi_{a},\phi_{b}$ fields computed in
Eq.(\ref{eq:intwgt}), together with the portion of the
multiboson energy depending on the neighboring multiboson fields
$\phi^{\prime}$ (which also depends on particle location only
through the dielectric field).
\end{enumerate}
     
A decision on a final location for the particle (either site $a$ or
site $b$) is now made on the basis of the relative weight determined
from the product of the three factors described above.  Once a
particle location has been decided, a new set of multiboson fields on
sites $a$ and $b$ are calculated by heat-bath by determining $\Psi$
according to the Gaussian Boltzmann weight $e^{-\Psi M\Psi}$, and
then setting
\[  \Phi= \left( \begin{array}{c}
  \phi_{a} \\ \phi_{b}    
\end{array} \right) = \Psi - \frac{1}{2}M^{-1}V \]     

\section{Simulation Results for the Coupled Heat-Bath Method}
\label{sec:results}

\begin{figure}
\centerline{\includegraphics[width=5in]{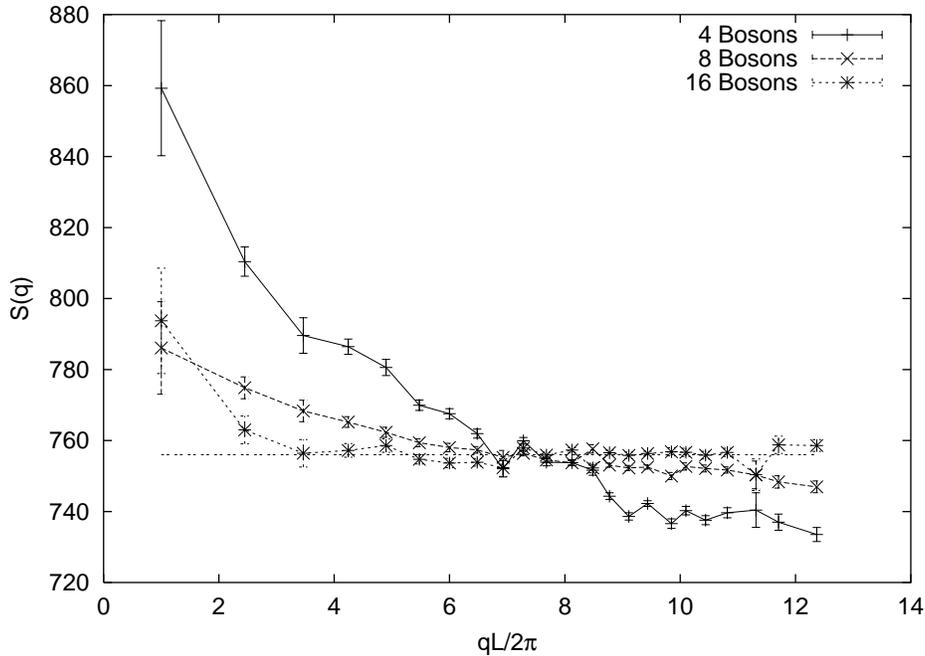}}
\caption{Structure factor for neutral system calculated using the
  coupled algorithm.  Results are shown for 4, 8, and 16 bosons.  160,000,
  400,000 and 800,000 measurement sweeps were used for the runs with
  4, 8, and 16 bosons respectively. For clarity, only every fifth
 wavenumber $q$ is plotted.}
\label{fig:structfact}
\end{figure}

\begin{figure}
\centerline{\includegraphics[width=5in]{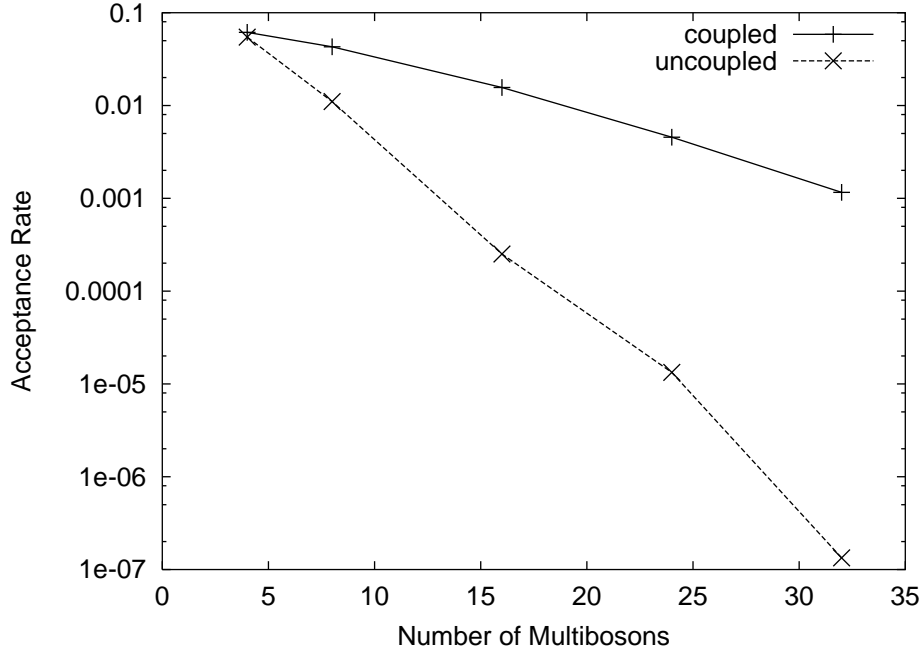}}
\caption{Acceptance rates for particle moves as a function of number
  of multibosons.  Shown for both earlier algorithm where the
  multiboson updates and particle updates are uncoupled and the
  improved coupled algorithm.}
\label{fig:accrates}
\end{figure}

We tested the efficiency of our improved simulation technique by
simulating a system of 1000 neutral particles on a 16x16x16 lattice as
studied in Ref.~\cite{ourmultibos}.  For the simulations of this paper
we have chosen a dielectric constant ratio between the particles and
the background medium to be 0.05 (in typical molecular dynamics
biophysical simulations, for example, the protein interior is given an
effective dielectric constant of $\approx$ 4 while the aqueous medium
is $\approx$ 80).  In particular, in the simulations the particles
have a dielectric constant of 0.05 and the background dielectric
constant is 1.0.  The dimensionless inverse temperature is 0.25, as in
the previous simulations of this system.

The relevant physical quantity for systems of this type is the
structure function defined as the Fourier transform $S(\vec{q})$ of
the coordinate space density-density correlation function
(Ref.~\cite{ourmultibos}). As we work on a spatial lattice, the
allowed momentum magnitude values $q\equiv|\vec{q}|$ are discrete: we
have chosen to average the structure factor $S(q)$ over degenerate
values of the magnitude of the Fourier momentum vector (on a 16x16x16
lattice, there are 115 nonzero distinct allowed values). This data set
is measured every 10 sweeps, and error bars extracted at the end of
the runs by blocking the measurements in order to take
autocorrelations into effect. This allows a measurement of a
chi-squared goodness-of-fit of the final averaged data to the exact
answer, which for neutral dielectric particles is simply a flat
structure function
\begin{eqnarray*}
\label{eq:exactstruc}
   S(q) = N\left(1-\frac{N-1}{V-1}\right),\;\;\;\; q\neq 0
\end{eqnarray*}
where $N$ is the number of particles and $V$ the number of spatial
points on the lattice.

As emphasized in Section 2, the choice of the number $N_B$ of
multiboson fields introduced to model the dielectric contribution to
the partition function is dictated by the need to properly describe
the low (infrared) part of the spectrum of the Poisson operator
$\cal{M}$, which contains eigenvalues of order $1/L^2$ on $L$x$L$x$L$
lattices. Too small a choice for $N_B$ can be expected to lead to
systematic deviations in the structure factor of
Eq.~\ref{eq:exactstruc} for small momenta $q$. On the other hand, too
large a choice leads to rapidly falling acceptance rates for
multiboson updates and is clearly more computationally costly per
Monte Carlo update step. The systematic deviation in the structure
factor resulting from too small a choice for $N_B$ is shown in
Fig.~\ref{fig:structfact}, where we compare the structure factors on a
16x16x16 lattice for $N_B=$4,8,16. It is clear that for this problem,
4 multiboson fields are inadequate, giving a distinct curvature to the
structure factor, and a large chi-squared per degree of freedom of
36.4, whereas the chi-squared per d.o.f.  for the simulation with 16
multiboson fields is only 1.94.

Depending on the physical parameter regime, and the size of lattice
used, it will evidently be necessary to increase $N_B$ to achieve
adequate accuracy, at least for long-range features of the
physics. Unfortunately, the update procedure introduced in
Ref. \cite{ourmultibos} suffers from the drawback that, for fixed
physical parameters (lattice size, dielectric constant, etc), the
acceptance rates for multiboson updates decreases exponentially
rapidly with $N_B$. This makes the uncoupled multiboson method
impractical for much larger lattices than those studied in our
previous work. In Fig.~\ref{fig:accrates} we compare the acceptance
rates using the original update procedure, in which particle moves
were decoupled from updates of the multiboson fields, with the
acceptance rates using the new procedure described in Section 3, in
which particle moves are coupled to readjustments in the multiboson
field. For larger values of $N_B$ it is clear that the acceptance
rates are increased by orders of magnitude (e.g. by about $10^{4}$ for
$N_B$=32).

Although the new, coupled procedure is more computationally intensive,
it does not require significantly more computational time per Monte
Carlo sweep.  We observe that for a run with the parameters described
above with 16 bosons, the computational time per Monte Carlo sweep
using the coupled update method is only about 13\% greater than in the
original uncoupled algorithm.  This additional computational cost is
insignificant given the great increase in particle move acceptances
that the coupled method provides.

\section{Conclusion}
\label{sec:conclusion}

By using an update method where the proposed particle moves include a
relaxation of the multiboson field to adjust to the change in particle
location we have been able to dramatically increase the acceptance
rates for larger systems that require a larger number $N_B$ of
multiboson fields to accurately model the long wavelength physics of
the system. For both the original uncoupled and the improved coupled
update algorithms, and for the physical parameters used here, the
dependence of the acceptance rate on $N_B$ is exponential
($\exp{(-KN_B)}$).  The coefficient in the exponent, $K$, is
$\simeq 0.13$ for the coupled algorithm and $\simeq 0.46$ for the
original method, amounting to a difference in acceptance rates of 4
orders of magnitude for a simulation with 32 multiboson fields.  Tests
done with charged particles show similar improvements to the acceptance
rates, although for some parameter regions the electric field must
also be coupled into the particle moves to obtain reasonable
acceptance rates \cite{ourpaper}.


\begin{acknowledgments}

The work of A.~Duncan was
supported in part by NSF grant PHY0244599. 
The work of R.D.~Sedgewick  was supported in part by the David
and Lucile Packard Foundation.
\end{acknowledgments}


\begin{thebibliography}{99}
\bibitem{ourmultibos} A. Duncan, R.D. Sedgewick and R.D. Coalson,
  Phys.\ Rev.\ E \textbf{73}, 016705 (2006)
\bibitem{maggs:prl} A.C.~Maggs and V.~Rossetto, Phys.\ Rev.\ Lett.\ \textbf{88}
, 196402 (2002)
\bibitem{maggs:alone} A.C.~Maggs, J.\ Chem.\ Phys.\ \textbf{120}, 3108
  (2004)
\bibitem{maggs:worm} L.~Levrel, F.~Alet, J.~Rottler, and
A.C.~Maggs, Pramana \textbf{64}, 1001 (2005)
\bibitem{maggs:trail} L.~Levrel and A.C.~Maggs, \texttt{cond-mat/0503744} (2005)
\bibitem{ourpaper:fft} A.~Duncan, R.D.~Sedgewick, and R.D.~Coalson,
  \texttt{cond-mat/0508266} (to be published in Comp.\ Phys.\ Comm.)
\bibitem{ourpaper} A.~Duncan, R.D.~Sedgewick, and R.D.~Coalson, Phys.\ Rev.\ E \textbf{71}, 046702 (2005)
\bibitem{luescher} M. L\"uscher, Nucl.\ Phys.\ B \textbf{418}, 637 (1994)
\bibitem{alet:orig_worm} F.~Alet and E.~S\o rensen, Phys.\ Rev.\ E
  \textbf{67}, 015701 (2003)


\end{thebibliography}
\end{document}